\documentstyle[psfig,aps,floats,eqsecnum]{revtex}

\begin{document}
\title{Quantum mechanics of measurement}

\author{N. J. Cerf$^1$ and C. Adami$^{1,2}$}
\address{$^1$W. K. Kellogg Radiation Laboratory and 
         $^2$Computation and Neural Systems,\\
California Institute of Technology, Pasadena, California 91125}

\date{Received 6 May 1996, revised January 1997}

\draft
\maketitle

\begin{abstract}
An analysis of quantum measurement is presented that relies on an
information-theoretic description of quantum entanglement.  In a
consistent quantum information theory of entanglement, entropies
(uncertainties) {\em conditional} on measurement outcomes can be {\em
negative}, implying that measurement can be described via unitary,
entropy-conserving, interactions, while still producing randomness in
a measurement device. In such a framework, quantum measurement is not 
accompanied by a wave-function collapse, or a quantum jump. The theory is
applied to the measurement of incompatible variables, giving rise to a 
stronger entropic uncertainty relation than heretofore known. 
It is also applied to standard quantum measurement situations such as
the Stern-Gerlach and double-slit experiments to illustrate how 
randomness, inherent in the conventional quantum probabilities, 
arises in a unitary framework. Finally, the present view clarifies the 
relationship between classical and quantum concepts.
\end{abstract}

\pacs{PACS numbers: 03.65.Bz,89.70.+c
              \hfill KRL preprint MAP-198}
\bigskip\bigskip
\centerline{ Submitted to Physical Review A.}

\narrowtext
\twocolumn

\section{Introduction}
\label{sect_introduction}

For seventy years it has remained a mystery how quantum measurement
can be probabilistic in nature, and thus be accompanied by the
creation of randomness or uncertainty, while at the same time being
described by unitary evolution. This apparent contradiction has cast
serious doubts on the very foundations of quantum
mechanics. Meanwhile, the equations and predictions of the purportedly
flawed theory enjoy unbridled, unequivocal success. In this paper, we
present an information-theoretic description of quantum measurement
which sheds new light on this long-standing question.
This description, in terms of the quantum information theory (QIT)
introduced by us recently~\cite{bib_neginfo,bib_qit}, implies that the
(conditional) quantum entropy of an entangled subsystem can be {\em
negative}, in contrast to its classical counterpart.  As we outline
below, this allows for the creation of entropy in the measurement
device which is counterbalanced by the negative entropy of the quantum
system itself, resulting in the conservation of entropy in the
measurement process. Consequently, the probabilistic nature of quantum
mechanics can be shown to follow from a completely consistent,
unitary, description of measurement.
\par

Our model does not require the quantum
system to be coupled to a macroscopic---uncontrollable---environment,
and is therefore distinct from the environment-induced decoherence
model, one of the prevalent contemporary views of quantum
measurement~\cite{bib_zurek}.  As shown below, the view advocated here
only insists on the ``self-consistency'' of the measurement device
while abandoning the 100\% correlation between the device and the
measured quantum system which is a cornerstone of decoherence models.
Also, while 
the information-theoretic interpretation suggests that the universe exists 
in a {\em superposition} of quantum states, all of quantum phenomenology 
is explained armed only with one, rather than many, such universes. 
\par

A more detailed investigation of the measurement process reveals 
that the collapse of the wavefunction is an illusion, brought about by 
the observation of part of a composite system that is
quantum entangled and thus
{\em inseparable}. Rather than collapsing, the wavefunction of 
a measured system becomes entangled with the wavefunction of the measurement 
device. If prepared in a superposition of 
eigenstates, the measured system is {\em not} reduced to one of 
its eigenstates. In other words, a quantum jump does {\em not} occur. 
That this must be the case has of course been 
suspected for a long time, and it certainly is implicit in the quantum
eraser experiments on which we shall comment below. Here we show that 
this feature emerges naturally if quantum entanglement is properly described
in the language of QIT. Furthermore, due to the absence of a collapse
of the wavefunction, our unitary description implies that quantum measurement 
is inherently {\em reversible}, overturning the common view. However,
in an experiment where quantum entanglement is transferred to a macroscopic
``pointer'' variable (as is essential for classical observers)
the reversibility is obscured by the {\em practical} impossibility of keeping
track of all the atoms involved in the unitary transformation, rendering
the measurement as irreversible as the thermodynamics of 
gases\footnote{This last conclusion is reached in
environment-induced decoherence models as well, since there is no qualitative
difference between an environment and a large number of degrees of freedom
belonging to a macroscopic measurement device.}. 
Thus, as suggested earlier by Peres~\cite{bib_peres74},
the apparent irreversibility of quantum measurement can be understood
entirely in classical terms. 
\par

In the next section, we briefly review the current state of quantum
measurement theory, with emphasis on the standard von Neumann theory 
of measurement.
In Section \ref{sect_qit}, we outline those features 
of the quantum information theory introduced in 
\cite{bib_neginfo,bib_qit} which apply to quantum measurement, and
point out the singular importance of negative entropy in quantum entanglement. 
We also focus on the relation between entanglement and inseparability
in this theory.
In Section \ref{sect_measprocess} we then proceed with a microscopic 
description of the unitary
physical measurement process as anticipated by von Neumann, but properly 
interpreted within QIT. 
We focus on the measurement of incompatible variables in
Section \ref{sect_incompatible} and show how one of the milestones of 
quantum physics, the 
uncertainty relation, emerges naturally from our construction. Alternatively,
this Section can be read as describing unitary quantum measurement 
more formally, implying some of the well-known relations of conventional 
quantum mechanics. 
Section \ref{sect_interpretation} discusses new insights into
the interpretation of quantum mechanics brought about by 
this information-theoretic analysis.
There, we investigate the relationship between classical and quantum variables
and propose a simple resolution to the ``Schr\"odinger-cat'' paradox.
Also, we comment on the origin of the complementarity
principle and the duality between waves and particles.
We offer our conclusions in 
Section \ref{sect_conclusion}. Finally, Appendix \ref{sect_application} 
illustrates the interpretation of standard experiments of quantum 
mechanics within our framework. There, we consider the basic 
Stern-Gerlach setup and ``quantum erasure'' in the standard double-slit 
experiment.

\section{Theory of measurement}
\label{sect_theorymeas}

The theory of measurement occupies a central role in quantum physics and
has undergone a number of conceptual revolutions. Those range from
the probabilistic interpretation of quantum mechanics by Born and
the Copenhagen interpretation championed by Bohr (see e.g. \cite{bib_wheeler}),
over
von Neumann's seminal contribution in the ``{\it Grundlagen}''
\cite{bib_vonneumann} to more modern interpretations such as Everett's
\cite{bib_everett,bib_dewitt}, Cramer's~\cite{bib_cramer},
and Zurek's~\cite{bib_zurek}.
\par

Central to all these 
treatments is the problem of the collapse of the wavefunction, or state 
vector. To illustrate this process, consider for example the measurement of 
an electron, described by the wavefunction $\Psi(q)$
where $q$ is the coordinate of the electron. Further, let the measurement 
device be characterized initially by its
eigenfunction $\phi_0(\xi)$, where $\xi$ may summarize the 
coordinates of the device. Before measurement, i.e., before the 
electron interacts with the measurement device, the system is described 
by the wavefunction 
\begin{equation}
\Psi(q)\phi_0(\xi)\;.
\end{equation}
After the interaction, the wavefunction is a superposition
of the eigenfunctions of electron and measurement device
\begin{equation}
\sum_n \psi_n(q)\phi_n(\xi)\;. \label{sum}
\end{equation}
Following orthodox measurement theory, the classical nature of the 
measurement apparatus implies that after measurement
the ``pointer'' variable $\xi$ takes on a well-defined value at each point in
time; the wavefunction, as it turns out, is thus {\em not} given by the
entire sum in (\ref{sum}) but rather by the single term
\begin{equation}
 \psi_n(q)\phi_n(\xi)\;.\label{term}
\end{equation}
The wavefunction (\ref{sum}) is said to have collapsed to (\ref{term}).
\par

A cornerstone of the Copenhagen interpretation of measurement was precisely
this collapse, due to the interaction of a quantum object with
a macroscopic, {\em classical}, measurement device.  
The crucial step to describe the measurement process as an interaction of
two {\em quantum} systems [as is implicit in (\ref{sum})] was made 
by von Neumann~\cite{bib_vonneumann}, who recognized that an interaction 
between a classical 
and a quantum system cannot be part of a consistent quantum theory.
In his {\it Grundlagen},
he therefore proceeded to decompose the quantum measurement into {\em two}
fundamental stages. The first stage (termed ``von Neumann measurement'')
gives rise to the wavefunction (\ref{sum}). The second stage (which 
von Neumann termed ``observation'' of the measurement)
involves the collapse described above, i.e., the transition from
(\ref{sum}) to (\ref{term}). 
\par
We now proceed to describe the first stage in more detail.
For ease of notation, let us recast this problem
into the language of state vectors instead. The first stage 
involves the interaction of the quantum system $Q$
with the measurement device (or ``ancilla'') $A$. Both the quantum system
and the ancilla are fully determined by their state vector, yet, let us 
assume that the state of $Q$ (described by state vector $|x\rangle$)
is unknown whereas the state of 
the ancilla is prepared in a special state $|0\rangle$, say. The state 
vector of the combined system $|QA\rangle$ before measurement then is
\begin{equation}
|\Psi_{t=0}\rangle = |x\rangle|0\rangle \equiv |x,0\rangle\;.
\end{equation}
The von Neumann measurement is described by the unitary evolution of $QA$
via the interaction Hamiltonian
\begin{equation}
\hat H = -\hat X_Q \hat P_A \;,\label{ham}
\end{equation}
operating on the product space of $Q$ and $A$. Here, $\hat X_Q$ is the 
observable to be measured, and  
$\hat P_A$ the operator  {\it conjugate} to the
degree of freedom of $A$ that will reflect the result of the measurement.
We now obtain for the state vector $|QA\rangle$ 
after measurement (e.g.\ at $t=1$, putting $\hbar=1$) 
\begin{equation}
|\Psi_{t=1}\rangle=e^{i\hat X_Q \hat P_A}|x,0\rangle =
 e^{ix\hat P_A}|x,0\rangle = |x,x\rangle \;. \label{entang}
\end{equation}
Thus, the pointer in $A$ that previously pointed to zero now also points to 
the position $x$ that $Q$ is in. According to von Neumann, this simple 
operation reflects the {\em correlation} between $Q$ and $A$ introduced
by the measurement. In general, this unitary
operation rather introduces {\em entanglement}, which is beyond the 
classical concept of correlations. In fact, the creation of entanglement in a 
von Neumann measurement\footnote{A 
general measurement can be described using a positive-operator-valued
measure (POVM), based on the decomposition of the identity operator
into positive operators on the Hilbert space~\cite{bib_peres}. The von Neumann
measurement is a special case in which the positive
operators are the orthogonal projection operators $|X_Q\rangle \langle X_Q|$
(which sum to identity because of the closure relation). The restriction to a 
simple von Neumann measurement, however, is sufficient for our purposes
since a POVM can always be described as a von Neumann measurement
in an extended Hilbert space.}
is {\em generic}. This is illustrated for typical measurement situations
in Appendix \ref{sect_application}.

\par
The second stage in von Neumann's theory of measurement, 
the {\em observation} of the pointer variable by
a conscious observer (or a mechanical device with memory), is the
key problem of measurement theory and the central object of this paper.
Historically, this conundrum is usually couched 
into the question: ``At what point does the possibility of an outcome
change into actuality?'' 
In the interpretation of this stage, von Neumann finally conceded to Bohr,
who maintained that the ``observing'' operation (stage two), now 
distinct from the ``measuring'' process (stage one), is {\it irreversible}
and {\it non-causal}. At first glance, there appears to be no escape 
from this conclusion, as a pure state (a superposition) seems to evolve 
into a mixed state (describing all possible outcomes), a process that cannot 
be described by a unitary operation. This becomes 
more evident if we apply the unitary operation described above to 
an initial quantum state which is in a quantum superposition:
\begin{equation}
|\Psi_{t=0}\rangle = |x+y,0\rangle\;.
\end{equation}
Then, the linearity of quantum mechanics implies that
\begin{equation}
|\Psi_{t=1}\rangle = e^{i\hat X_Q\hat P_M}
\biggl(|x,0\rangle+|y,0\rangle\biggr)=
|x,x\rangle+|y,y\rangle\label{entang1}
\end{equation}
which is still a pure state. However, it does not reflect classical
correlations between $Q$ and $A$ (as would the state $|x+y,x+y\rangle$) but
rather {\em quantum entanglement}. This realization is the 
content of the celebrated quantum non-cloning theorem~\cite{bib_noncloning}.
Just like the wavefunction (\ref{sum}), the state vector (\ref{entang1})
cannot describe the result of the observation of the pointer, as the 
pointer is classical and takes on definite values. Thus, a measurement 
will reveal $A$ to be in the state $|x\rangle$ {\em or} $|y\rangle$, the sum
(\ref{entang1}) will appear to have collapsed, and a ``completely 
known'' (fully described) quantum object seems to have evolved into one of 
several possible outcomes. 
This recurrent problem forced von Neumann to introduce a process
{\em different} from unitary evolution to describe the second
stage in quantum measurement, the {\em observation} of $A$ in the entangled 
system $QA$. While he showed that the boundary between the observed 
system $QA$
and the observer can be placed arbitrarily, he still concluded 
that ``observation'' must ultimately take place. Reluctantly, he 
suggested that the collapse of the wavepacket had to occur
in the observer's brain, thereby allowing the concept of consciousness 
to enter in his description of 
measurement~\cite{bib_vonneumann,bib_london,bib_wigner}.
\par
To this date, there is no unanimous agreement on a solution to this
problem. A promising attempt at
unraveling the mystery was presented by Everett~\cite{bib_everett}. In his 
interpretation, measurement is described exactly as outlined above, 
only the second stage never takes place. Rather, the different terms 
in the sum (\ref{sum}) or (\ref{entang1}) are interpreted as the 
``records'' of (conscious or mechanical) observers, each recording
possible versions of reality, while only one particular term is 
available for one observer in a particular instantiation. The sum has 
been interpreted by DeWitt~\cite{bib_dewitt} as the wavefunction of a 
universe constantly branching at each quantum event. 
While internally consistent, the Everett--DeWitt interpretation suffers 
from the burden of unprovable {\it ad hoc} assumptions. 
Interesting from the point of view advanced here are the formulations
of Peres~\cite{bib_peres74} and Zurek~\cite{bib_zurek},
generally referred to as environment-induced decoherence models.
In their approach, mixed states are obtained
from pure states by tracing over either the measurement apparatus
(for example because it has many uncontrollable degrees of freedom)
or a macroscopic environment (which absorbs the quantum phases because
it involves enormously numerous random degrees of freedom).
The underlying idea thus is that the loss of information in a macroscopic
system is responsible for the creation of entropy in a measurement.
While accounting for the apparently irreversible character of
quantum measurement, this approach does not address the issue of the collapse,
nor does it provide a satisfying explanation for the Schr\"odinger-cat
paradox (see, e.g., \cite{bib_DNA}). Another interesting attempt
is due to Cramer~\cite{bib_cramer}, who invokes the exchange of
retarded and advanced waves between elements of a measurement situation
in the second stage of measurement. 
The difficulty to describe quantum measurement as unitary evolution
is affecting areas of physics as diverse as black holes and quantum optics. 
Attempts at tackling the problem range
from giving up unitarity in quantum mechanics to understand
the production of entropy in Hawking radiation~\cite{bib_nonliouville},
to describing quantum decoherence via a non-Liouvillian
equation~\cite{bib_decoherence}.
Most recently, it was suggested that using DNA as a 
microscopic measuring device~\cite{bib_DNA} (to record the absorption
of ultraviolet photons) would reveal that 
``[...] even the most prominent 
nonorthodox models of quantum mechanics have nontrivial difficulties'' if 
no essential role is ascribed to a conscious observer!

\par 
Historically, it appears that the failure to understand von Neumann's
second stage is rooted in a misunderstanding of the correlations
introduced by the first stage. In fact, it was only three years {\em after} 
the appearance of the {\it Grundlagen} that Einstein, Podolsky, and Rosen 
(EPR)~\cite{bib_epr} pointed out the peculiarities of a wavefunction 
such as (\ref{entang1}), now known
as the wavefunction of an EPR entangled state. As we shall see in the 
next Section, correlations inherent in such a state cannot be understood 
via classical concepts, as the state so created is not separable.  
The observation of only a {\em part} of such a system effects the 
appearance of probabilities (in a subsystem) when in fact none such are 
present (in the combined system). The second
stage of measurement can be understood {\em without} recourse to 
non-unitary time-evolution or the intervention of consciousness, 
within the language of the quantum
information theory introduced recently~\cite{bib_neginfo,bib_qit}. 
\par

\section{Quantum information theory}
\label{sect_qit}

In the standard information theoretical treatment of quantum measurement, 
classical (Shannon) information theory~\cite{bib_shannon}
is applied to probabilities derived from quantum mechanics.
More precisely, the quantum probabilities
of the different outcomes of the measurement of a quantum state are 
used to calculate the tradeoff between entropy and information that 
accompanies the measurement~\cite{bib_conventional}.
However, this treatment is incomplete, as
the quantum probabilities entering Shannon theory are devoid of the 
phase information which characterizes quantum mechanical superpositions.
To be consistent, quantum information theory needs to be based on density 
matrices only, rather than on probability distributions. 
\par

Let us summarize the unified information-theoretical description 
of correlation and entanglement that was introduced
in Ref.~\cite{bib_neginfo,bib_qit}. This theory
parallels classical (Shannon) information theory, but extends it to
the quantum regime. A quantum system $A$, described by a density
matrix $\rho_A$, has von Neumann entropy 
\begin{equation}
S(A)= - {\rm Tr}_A [\rho_A \log \rho_A]
\end{equation}
where ${\rm Tr}_A$
denotes the trace over the degrees of freedom associated with $A$.
If $\rho_A$ is expressed in a diagonal basis, i.e.,
$\rho_A=\sum_a p(a) |a\rangle \langle a|$,
the von Neumann entropy is equal to the classical (Shannon-Boltzmann-Gibbs)
entropy 
\begin{equation}
H(A)=-\sum_a p(a) \log p(a)\;.
\end{equation}
An important property of the von Neumann entropy $S(A)$ is that it
remains constant when the system $A$ undergoes a unitary transformation.
This is analogous to the Boltzmann entropy remaining constant
under a reversible transformation in classical thermodynamics. As quantum
mechanics only allows unitary time-evolution, the von Neumann entropy of
any isolated system remains constant in time. 
\par

\begin{figure}
\caption{ (a) General entropy diagram for a quantum composite system $AB$.
(b) Entropy diagrams for three cases of two spin-1/2 particles:
(I) independent, (II) classically correlated,
and (III) quantum EPR-entangled.  \label{fig1} }
\vskip 0.5cm
\par
\centerline{\psfig{figure=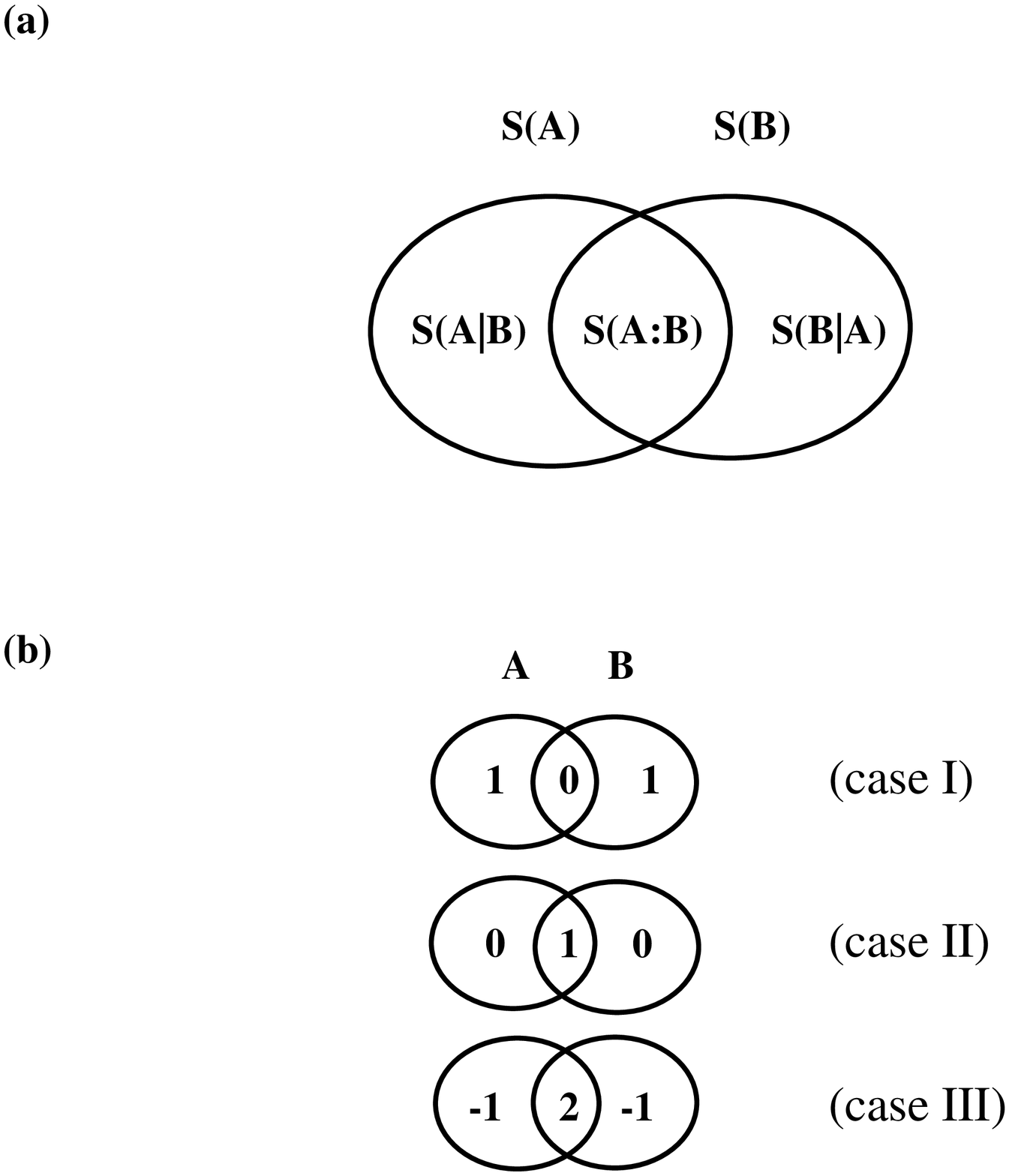,width=2.70in,angle=0}}
\par
\end{figure}

The substitution of probabilities (in classical information theory) 
by density matrices (in quantum information theory)
becomes crucial when considering composite systems, such as
a bipartite system $AB$.
Indeed, the density matrix $\rho_{AB}$ of the entire system  
can in general not be written as a diagonal matrix, if changes of 
basis are performed on the variables associated to $A$ and $B$ 
{\em separately}.
(Of course, $\rho_{AB}$ can always be diagonalized by applying a
change of variables to a joint basis.) The composite system $AB$
is associated with a von Neumann entropy
\begin{equation}
S(AB)=-{\rm Tr}_{AB} [\rho_{AB} \log \rho_{AB}] 
\end{equation}
Now, in order to analyze a measurement situation, we need to consider
a {\em conditional} quantum entropy $S(A|B)$, which describes
the entropy of $A$ {\em knowing} $B$.
Let $S(A|B)$ therefore
denote the von Neumann entropy of $A$ {\em conditional} on $B$,
and be given by
\begin{equation}  \label{condentr}
S(A|B)=-{\rm Tr}_{AB} [\rho_{AB} \log \rho_{A|B}]
\end{equation}
with 
\begin{equation}
\rho_{A|B}=\lim_{n\to\infty}
\left[\rho_{AB}^{1/n} ({\bf 1}_A \otimes \rho_B)^{-1/n}\right]^n
\label{condmat}
\end{equation}
the ``conditional'' density matrix defined in \cite{bib_neginfo}. Here,
$\otimes$ stands for the tensor product in the joint Hilbert space and
$\rho_B={\rm Tr}_A [\rho_{AB}]$
denotes a ``marginal'' (or reduced) density matrix,
obtained by a partial trace over the 
variables associated with $A$ only. The conditional density matrix defined 
here is just the quantum analogue of the conditional probability 
$p(a|b)=p(a,b)/p(b)$
in classical information theory and reduces to it in a classical situation
(i.e., when the density matrix is diagonal). In the case that $\rho_{AB}$ and
$({\bf 1}_A \otimes \rho_B)$ commute,
Eq.~(\ref{condmat}) simply reduces to
\begin{equation}
\rho_{A|B}=\rho_{AB}({\bf 1}_A \otimes \rho_B)^{-1}\;.
\end{equation}
Using Eqs. (\ref{condentr}) and (\ref{condmat}),
it can be checked that the total entropy decomposes as
\begin{equation}
S(AB)=S(A)+S(B|A) = S(B) + S(A|B)\;, \label{bayes}
\end{equation}
in perfect analogy with
the equations relating classical entropies.
We also define a quantum {\em mutual} entropy
\begin{equation}  \label{mutentr}
S(A{\rm :}B)= - {\rm Tr}_{AB} [\rho_{AB} \log \rho_{A:B}]
\end{equation}
with 
\begin{equation}
\rho_{A:B}=\lim_{n\to\infty}
\left[ (\rho_A \otimes \rho_B)^{1/n} \rho_{AB}^{-1/n} \right]^n
 \;, \label{mutmat}
\end{equation}
which reduces to
\begin{equation}
\rho_{A:B}=(\rho_A \otimes \rho_B) \rho_{AB}^{-1}
\end{equation}
for commuting matrices.
Using Eqs. (\ref{mutentr}) and (\ref{mutmat}),
the quantum mutual entropy can be written as  
\begin{equation}
S(A{\rm :}B)=S(A)+S(B)-S(AB) \label{mutent}
\end{equation}
and is interpreted as the ``shared'' entropy between
$A$ and $B$. Eqs. (\ref{bayes}) and (\ref{mutent}) precisely parallel the 
classical relations, and validate the definitions (\ref{condmat}) and
(\ref{mutmat}).
The relations between $S(A)$, $S(B)$, $S(AB)$, $S(A|B)$, $S(B|A)$,
and $S(A{\rm:}B)$ are conveniently summarized by 
a Venn-like entropy {\em diagram}, as shown in Fig.~\ref{fig1}a.
\par

As mentioned earlier, in spite of the apparent similarity between
the quantum definitions for $S(A|B)$ or $S(A{\rm:}B)$ and their
classical counterparts, dealing with matrices (rather than scalars)
opens up a quantum realm for information theory that is inaccessible
to classical physics.
The crucial point is that, while a conditional probability
is a probability distribution ({\it i.e.}, $0 \le p(a|b) \le 1$),
its quantum analogue $\rho_{A|B}$ is {\it not} a density matrix.
In general, $\rho_{A|B}$ is a positive Hermitian matrix in the joint
Hilbert space, but it can have eigenvalues exceeding one, and consequently,
the associated conditional entropy $S(A|B)$ can be negative.
In classical information theory, a conditional entropy $H(A|B)$ is
always non-negative. This is in agreement with common sense, since the
classical entropy of a composite system $AB$ cannot be lower than
the entropy of any subsystem $A$ or $B$. More precisely, for classical
entropies, we have the
basic inequality,
\begin{equation}  \label{eq_bound}
\max[H(A),H(B)] \le H(AB) \le H(A)+H(B) 
\end{equation}
where the upper bound is reached for independent subsystems, while the
lower bound corresponds to maximally correlated subsystems and
implies $H(A|B)\ge 0$, $H(B|A)\ge 0$. In contrast, the equivalent
inequality (due to Araki and Lieb~\cite{bib_araki})
for quantum entropies becomes
\begin{equation}
|S(A)-S(B)| \le S(AB) \le S(A)+S(B)
\end{equation}
where the lower bound can be {\it lower} than the classical one, implying
that $S(A|B)$ or $S(B|A)$ can be negative. This well-known non-monotonicity
of quantum entropies follows naturally in our matrix-based
formalism from the fact that $\rho_{A|B}$ can have eigenvalues larger
than one. The situation where $S(A)>S(AB)$ or $S(B)>S(AB)$ occurs 
in the case of quantum entanglement.

As an illustration, it is instructive to consider three simple cases
of two spin-1/2 particles with entropy\footnote{Throughout this
paper we take logarithms to the base two, such that entropies are 
expressed in {\em bits}.}$S(A)=S(B)=1$.
In our first case I, let the particles be independent, each one being 
described by the density matrix
\begin{equation}
\rho_A=\rho_B=\frac12\left(|\uparrow \rangle \langle \uparrow| + 
|\downarrow \rangle \langle \downarrow|\right)
\end{equation}
Then, the entire system has $\rho_{AB}=\rho_A \otimes \rho_B$,
so that the total entropy is $S(AB)=2$, while each system
carries one bit of entropy (see Fig.~\ref{fig1}b). Also,
we have $\rho_{A|B}=\rho_A \otimes {\bf 1}_B$ and 
$\rho_{B|A}={\bf 1}_A \otimes \rho_B$, implying that
$S(A|B)=S(A)$ and $S(B|A)=S(B)$. In our next case II, 
let the two particles be fully (classically) correlated, so that
\begin{equation}
\rho_{AB}=\frac12(|\uparrow\uparrow \rangle \langle \uparrow\uparrow| + 
|\downarrow\downarrow \rangle \langle \downarrow\downarrow|)\;.
\end{equation}
This is a uniform mixture, with the two particles always in the same state
(i.e., classically correlated). The respective entropies are shown in
Fig.~\ref{fig1}b. Our last 
case III is quantum entanglement, and corresponds physically to
the situation which appears when a singlet state is created 
by the decay of a spin-0 particle into two spin-1/2 particles (creating an
``EPR-pair''). Such a system is described by the EPR 
wave function\footnote{
The state in (\ref{bell}) is in fact one of the {\em Bell} states, which are a 
generalization of the EPR state.}
\begin{equation}       \label{bell}
|\psi_{AB}\rangle =\frac1{\sqrt{2}}(|\uparrow\uparrow \rangle 
+ |\downarrow\downarrow \rangle)\;.
\end{equation}
Here, $\rho_{AB}=|\psi_{AB}\rangle \langle \psi_{AB}|$, so that we have
$S(AB)=0$, as expected for a pure quantum state. By taking a partial
trace of $\rho_{AB}$, we see that both subsystems $A$ and $B$ are
in a mixed state
\begin{equation}
\rho_A=\rho_B=\frac12(|\uparrow \rangle \langle \uparrow| 
+ |\downarrow\rangle\langle\downarrow|)\;,
\end{equation}
as in cases I and II. Such mixed states have positive entropy, yet, 
the combined entropy is zero in this case. Then,
the conditional entropies are forced to be {\em negative}, 
$S(A|B)=S(B|A)=-1$, whereas the mutual entropy $S(A:B)=2$
(this is illustrated in Fig.~\ref{fig1}b). This can be verified by 
straightforward evaluation. In general, conditional entropies are
negative for {\em any} isolated ($S=0$) entangled quantum system.
Note further that the EPR entanglement  
constraint $[S(AB)=0]$ for an EPR pair arises 
from the fact that it is created via a unitary transformation
from a system initially in a zero entropy pure state (the decay of the 
spin-0 particle). This constraint implies that only one of the three entropies
$S(A|B)$, $S(B|A)$, and $S(A:B)$, is an independent variable. In other words,
the entropy diagram of {\em any} pure entangled bipartite system
can only be a multiple of that of case III in Fig.~\ref{fig1}b. 
This situation violates the classical inequalities [Eq.~(\ref{eq_bound})]
that relate Shannon entropies, and therefore corresponds to a purely quantum 
situation, while cases I and II  
are classically allowed~\cite{bib_neginfo,bib_qit}.
In this sense, the matrix-based framework presented above must be
seen as an extension of Shannon theory: it describes all the situations
allowed classically (from case I to case II), but extends to entanglement
(case III).
\par

The appearance of ``unclassical'' ($>1$) eigenvalues in the conditional
density matrix of entangled states 
can be related to quantum non-separability and the
violation of entropic Bell inequalities, 
as shown elsewhere~\cite{bib_bellpaper}.
As far as the separability of a {\em pure} state is concerned,
it is straighforward to check that
the non-negativity of the conditional
entropy is a necessary and sufficient condition for separability.
The separability of {\em mixed} states, on the other hand, presents
a more difficult problem.
First, the concavity of $S(A|B)$ in $\rho_{AB}$, a property related
to strong subadditivity of quantum entropies, implies that any separable
state~\cite{bib_werner}
\begin{equation}  \label{separable}
\rho_{AB} = \sum_k w_k \, \rho_A^{(k)} \otimes \rho_B^{(k)} 
\qquad ({\rm with~}\sum_k w_k=1)
\end{equation}
is associated with a non-negative conditional entropy $S(A|B)$.
(The converse is not true.)
Indeed, each product component $\rho_A^{(k)} \otimes \rho_B^{(k)}$
of a separable state is associated with the conditional density matrix
\begin{equation}
\rho_{A|B}^{(k)}=\rho_A^{(k)} \otimes {\bf 1}_B
\end{equation}
so that we have
\begin{equation}
S(A|B) \ge \sum_k w_k S(\rho_A^{(k)}) \ge 0  \;.
\end{equation}
This shows that the non-negativity of conditional entropies is a
{\it necessary} condition for separability. This condition is shown
to be equivalent to the non-violation of entropic Bell inequalities
in Ref.~\cite{bib_bellpaper}.
Secondly, it is easy to check from Eq.~(\ref{condentr}) that,
if $S(A|B)$ is negative, $\rho_{A|B}$ must admit at least one
``non-classical'' eigenvalue, {\it i.e.}, an eigenvalue exceeding
one. This results from the fact that ${\rm Tr}(\rho \sigma)\ge 0$
if $\rho$ and $\sigma$ are positive (Hermitian) matrices.
We have checked that {\it all} the eigenvalues of $\rho_{A|B}$
and $\rho_{B|A}$ are $\le 1$ for randomly generated
separable density matrices [of the form Eq. (\ref{separable})],
which suggests the conjecture that the ``classicality'' of the spectrum
of $\rho_{A|B}$ is a (strong) necessary condition
for separability\footnote{Note that the spectrum of $\rho_{A|B}$ and
$\rho_{B|A}$ is invariant under local transformations of the
form $U_A \otimes U_B$.}. 
\par

For example, this criterion can be applied to two spin-1/2
particles in a Werner state, that is a mixture of a singlet fraction $x$
and a random fraction $(1-x)$, as recently examined by
Peres~\cite{bib_peressepar}. The density matrix of this state
is given by
\begin{equation}
\rho_{AB}=\left( \begin{array}{cccc}
{(1-x) / 4} & 0 & 0 & 0 \\
0 & {(1+x) / 4} & {-x / 2} & 0 \\
0 & {-x / 2} & {(1+x) / 4} & 0 \\
0 & 0 & 0 & {(1-x)/ 4} \end{array} \right)
\end{equation}
A simple calculation shows that $\rho_{A|B}$ admits three eigenvalues equal
to $(1-x)/2$, and a fourth equal to $(1+3x)/2$. The above separability
criterion is thus fulfilled when this fourth eigenvalue does not
exceed 1, that is for $x \le 1/3$. Therefore, for this particular case,
our condition simply reduces to Peres' condition based on the positivity
of the partial transpose of $\rho_{AB}$.\footnote{Peres' criterion
of separability~\cite{bib_peressepar}
is that none of the eigenvalues of the partial transpose of $\rho_{AB}$
is negative. For the Werner state, three eigenvalues are equal to 
$(1+x)/4$ and the fourth one is equal to $(1-3x)/4$. This lowest eigenvalue
is non-negative if $x\le 1/3$. Thus, expressing that these eigenvalues
are non-negative is simply equivalent to expressing that the eigenvalues
of $\rho_{A|B}$ do not exceed one.}
(It happens to be a sufficient condition for a $2\times 2$ Hilbert space.)
We have checked, however, that our criterion is distinct from Peres'
in general, opening the possibility that it could be a stronger
necessary (or perhaps sufficient) condition for separability in a Hilbert
space of arbitrary dimensions. Further work will be devoted to this question.
\par

\begin{figure}
\caption {Ternary entropy Venn-diagram for a general tripartite
system $ABC$. The component entropies are defined in the text.}
\label{fig_general3}
\vskip 0.25cm
\centerline{\psfig{figure=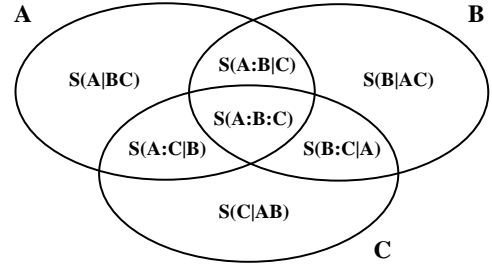,width=2.50in,angle=-90}}
\end{figure}

The description of quantum entanglement within this
information-theoretic framework turns out to be very
powerful when considering tripartite -- or more generally
multipartite-- quantum systems. Indeed, it is possible to extend
to the quantum regime the various classical entropies that are defined
in the Shannon information-theoretic treatment of a multipartite system.
This accounts for example for the emergence of classical correlation
from quantum entanglement in a tripartite (or larger) system.
Also, the quantum analogues of all the fundamental relations between
classical entropies (such as the chain rules for entropies and mutual
entropies) hold in quantum information theory and have the same
intuitive interpretation. 
Let us first consider a simple diagrammatic way of representing
quantum entropies involved in a tripartite system $ABC$, as shown
in Figure~\ref{fig_general3}.

The conditional entropies $S(A|BC)$,
$S(B|AC)$, and $S(C|AB)$ are a straightforward generalization
of conditional entropies in a bipartite system, that is
$S(A|BC)=S(ABC)-S(BC)$, etc. The entropies $S(A{\rm:}B|C)$,
$S(A{\rm:}C|B)$, and $S(B{\rm:}C|A)$ correspond to conditional mutual
entropies, {\it i.e.} the mutual entropy between two of the
subsystems when the third is known. In perfect analogy with the classical
definition, one can write, 
\begin{eqnarray}
S(A{\rm:}B|C) &=& S(A|C)-S(A|BC)  \nonumber \\
&=& S(AC)+S(BC)-S(C)-S(ABC)
\end{eqnarray}
which illustrates that the conditional mutual entropies
are always non-negative as a consequence of the strong subadditivity
property of quantum entropies.
The entropy in the center of the diagram is
a {\it ternary} mutual entropy, defined as
\begin{eqnarray} 
S(A{\rm:}B{\rm:}C)&=&S(A{\rm:}B)-S(A{\rm:}B|C)  \nonumber \\
&=&S(A)+S(B)+S(C)-S(AB) \nonumber \\
& & -S(AC)-S(BC)+S(ABC)
\end{eqnarray}
and corresponds to the entropy shared by the three subsystems
$A$, $B$, and $C$. Note that for any tripartite system
in a pure state, we have $S(AB)=S(C)$, $S(AC)=S(B)$, and $S(BC)=S(A)$,
so that the {\it ternary} mutual entropy vanishes.
More generally, for a multipartite system, relations between
quantum entropies can be written which parallel the classical
relations and have the same intuitive interpretation.
\par

\begin{figure}[t]
\caption{ (a) Ternary entropy diagram for a GHZ state (an ``EPR-triplet'').
(b) Entropy diagram for subsystem $AB$, unconditional on $C$.
The entropy of $C$ conditional on $AB$ is negative, and compensates
the positive entropy of $AB$ unconditional on $C$.  \label{fig2} }
\vskip 0.5cm
\par
\centerline{\psfig{figure=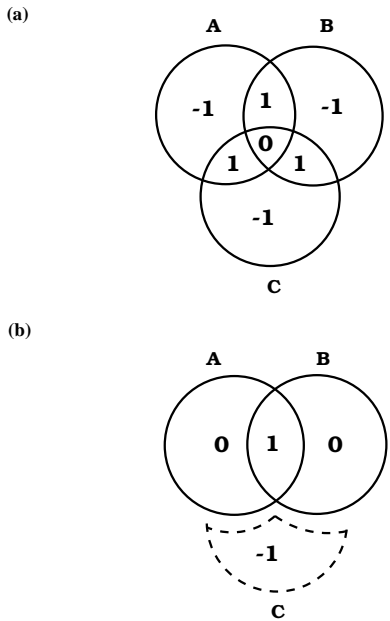,width=2.0in,angle=0}}
\end{figure}

As an illustration, let us consider a tripartite system $ABC$ in a
Greenberger-Horne-Zeilinger (GHZ) state
(which will become crucial in the quantum measurement process),
described by the wave function
\begin{equation}  
|\psi_{ABC}\rangle = \frac1{\sqrt2}(|\uparrow\uparrow\uparrow\rangle + 
|\downarrow\downarrow\downarrow\rangle)\;.
\end{equation}
As it is a pure state, its quantum entropy is $S(ABC)=0$.
When tracing over any degree of freedom 
(for instance the one associated with $C$), we obtain
\begin{equation}
\rho_{AB}=\frac12(|\uparrow\uparrow \rangle \langle \uparrow\uparrow| 
+ |\downarrow\downarrow \rangle \langle \downarrow\downarrow|)
\end{equation}
corresponding to a classically correlated system of type II 
(see Fig.~\ref{fig1}b). We thus find
$S(A)=S(B)=S(C)=S(AB)=S(AC)=S(BC)=1$, allowing us to
fill in the entropy diagram\footnote{The negative conditional
entropies in this diagram betray that 
this state is purely quantum, unobtainable in classical 
physics. As mentioned earlier, the fact that the {\em ternary} mutual
entropy $S(A{\rm:}B{\rm:}C)$ is zero is generic
of the description of any three-body system
in a pure state [it follows from the
constraint $S(ABC)=0$, i.e., that $ABC$ has been formed by applying
a unitary transformation on a pure state].} for the GHZ state
in Fig.~\ref{fig2}a.
The important feature of the GHZ state is that it entails quantum entanglement
between {\em any} part (e.g., $C$) and the rest of the system ($AB$). 
Even more important, ignoring (that is, tracing over) a part of 
it ($C$) creates {\it classical} correlation between the two remaining
parts ($A$ and $B$), as shown in Fig.~\ref{fig2}b. In other words, 
the subsystem 
$AB$ {\em unconditional} on $C$, i.e., without considering the
state of $C$, is indistinguishable from a type II system.
This property is {\em central} to the understanding of the quantum 
measurement process, and will be emphasized throughout the following section.
It is generalized without difficulty to the case of an ``EPR-nplet'':
\begin{equation}
|\psi\rangle=\frac1{\sqrt2}(|\uparrow\uparrow\cdots \uparrow\rangle 
+ |\downarrow\downarrow\cdots \downarrow\rangle)\;.
\end{equation}
Ignoring (tracing over) any degree of freedom creates classical 
correlations between all the remaining degrees of freedom.
\par

We can see now how an EPR entangled system (an EPR pair) plays a 
special role in quantum mechanics. The correlation between the 
elements of the pair [described by the mutual entropy
$S(A{\rm:}B)$] goes beyond anything classically achievable
(``super-correlation'').
A classical approach to understanding the correlations suggests 
that measuring half of an EPR pair {\em immediately} affects the
other half, which may be arbitrarily far away. Classical thinking of this 
sort applied to
an EPR pair is misleading, however. Indeed, a careful investigation of the 
information flow
in EPR pair experiments reveals that causality is never violated. In 
Ref.~\cite{bib_neginfo} we suggest that  
EPR pairs are better understood in terms of
qubit--antiqubit pairs, where the qubit (antiqubit) carries
plus (minus) one bit of information, and antiqubits are interpreted as 
qubits traveling {\em backwards} in time\footnote{The term {\em qubit} 
denotes the quantum unit of information, which is the quantum analog to the 
classical unit of information), see, e.g.~\cite{bib_bennett}.}.  
In anticipation of the discussion 
in the following section, let us mention (as advertised earlier) that the 
von Neumann measurement [see Eq.~(\ref{entang1})] creates just such
EPR entanglement (not classical correlation) between the quantum system 
and the measurement device. The key realization will be
that the quantum von Neumann entropy rather than
Shannon-Boltzmann-Gibbs entropy is in fact the {\em physical} 
entropy~\cite{bib_neginfo,bib_qit}. 
This explains the observation that entropy is created in the 
measurement of the spin of, say, an electron, in spite of the fact that
the von Neumann entropy is zero for a pure state, independently of the
choice of basis.
As we outline below, the apparent entropy created in a spin measurement 
(if the spin is not aligned with the measurement axis) is actually the 
quantum entropy of {\em part} of an entangled system, and is cancelled by 
the negative conditional entropy of the (non-observed) remainder.
\par

\section{Measurement process}
\label{sect_measprocess}

\subsection{Second stage: observation} 

We have now prepared the ground to understand von Neumann's second stage.
The crucial observation was touched upon briefly 
above: von Neumann entanglement (\ref{entang1}) creates 
{\em super-correlations} (a type III EPR-entangled state) between 
$Q$ (measured quantum system) and $A$ (ancilla), rather than
correlations. The system $QA$ thus created is inherently quantum, and cannot
reveal any classical information. To obtain the latter, we need to create
classical correlations between {\em part} of the EPR-pair $QA$ and 
{\em another} ancilla $A^\prime$, i.e., we need to 
observe the quantum observer. No new ingredients are needed for this.
Rather, we simply allow the EPR entangled system $QA$ to come into 
contact with a system $A^\prime$, 
building the system $QAA^\prime$. Subsequently, we apply a unitary 
transformation with an interaction Hamiltonian of the type (\ref{ham}),
only that now it is defined on the combined Hilbert space of $QA$ and 
$A^\prime$. Clearly, this is just a repetition of the first stage, but
now leading to a GHZ-like state\footnote{We dispense with normalizations.}
\begin{equation}
|QAA^\prime\rangle= |x,x,x\rangle + |y,y,y\rangle \;. \label{triplet}
\end{equation}
All operations have been unitary, and $QAA^\prime$ is described by the pure
state
\begin{equation}
\rho_{QAA^\prime}=|QAA^\prime\rangle\langle QAA^\prime| \;.
\end{equation} 
Experimentally, however, we are {\em only} interested in the correlations 
between $A$ and $A^\prime$, and {\em not} in correlations between
$A$ and $Q$ (which are unobservable anyway).
Luckily, there is no obstacle to obtaining such
classical (type II) correlations now (unlike in the case where only two
particles were quantum entangled). Indeed, it is now immediately obvious 
that when ignoring the quantum state $Q$ {\em itself}, as paradoxically
as it may appear at first sight, $A$ and $A^\prime$ find themselves 
classically correlated and in a {\em mixed} state:
\begin{equation}
\rho_{AA^\prime}={\rm Tr}_Q(\rho_{QAA^\prime})=|x,x\rangle\langle x,x|
+|y,y\rangle\langle y,y|\;.
\end{equation}
We will show that ignoring $Q$ turns out to be unavoidable when measuring
$Q$. This is the basic operation (ignoring part of an 
``EPR-nplet'') that was alluded to in the previous section, and which we will 
encounter again below. 
\par

In general, for the measurement of any quantum system in an 
$N$-dimensional discrete Hilbert space we obtain after tracing over $Q$
\begin{equation}
\rho_{AA^\prime}=\sum_{i=1}^N\,p_i\,|ii\rangle\langle ii| \label{traced}
\end{equation}
where the $p_i$ are the probabilities to find $A$ (or $A^\prime$) in
one of its eigenstates $|i\rangle$. This completes the second stage of 
the quantum measurement. A state was formed ($AA^\prime$) which 
{\em appears} to be mixed, 
\begin{equation}
S(AA^\prime) > 0  \;,
\end{equation}
while $A$, $A^\prime$ and $Q$ were pure to 
begin with. Yet, this mixed state is quantum entangled with $Q$, which carries
negative conditional entropy 
\begin{equation}
S(Q|AA^\prime) < 0
\end{equation}
such that the combined system $QAA^\prime$ is still pure:
\begin{equation}
S(QAA^\prime) = S(AA^\prime) + S(Q|AA^\prime) =0  \;.
\end{equation} 
Clearly therefore, a transition from a pure state to a mixed state 
(for the entire isolated system $QAA^\prime$) did
{\em not} take place, whereas the quantum probabilities
in the mixed state $AA^\prime$ correspond {\em precisely} to the square 
of the amplitudes of quantum mechanical measurement (see 
Section~\ref{sect_incompatible}). 
Quantum probabilities arise in unitary time development,
thanks to the negative entropy of the ``unobserved'' quantum system $Q$.
\par

Let us emphasize now the fact that this view of measurement implies that
conceptually {\em three} rather than just two systems must be involved.
The ``observation'' of the measurement is possible only when a third system 
$A^\prime$ (a quantum particle or set of particles
with a Hilbert space dimension 
at least equal to the dimension of the Hilbert space of $Q$)
interacts with $A$ (the ancilla which ``measured'' $Q$ through
von Neumann entanglement). Indeed, the classical intuition of measurement 
is built upon {\em correlations}, which can only emerge in the presence of a 
{\em third} system $A^\prime$.
The fact that $A^\prime$ need not be a microscopic
object is an issue which will become important when we will be concerned
with the {\em amplification} of the measurement. But, conceptually speaking,
it is enough to say that $A^\prime$ is a particle that ``observes'' 
the measurement made by $A$ on $Q$. Because classical observers are 
necessarily made out of a macroscopic number of
particles, it is {\em in practice} necessary to have a large number 
of correlated particles $A^\prime,A^{\prime\prime},\cdots$ in order to 
achieve a macroscopic measurement. However,
this is completely arbitrary: we may say that a measurement has been
performed as long as the result is recorded on any kind of storage 
device\footnote{This is the content of the so-called ``psychophysical 
parallelism'' hypothesis, that a measurement is achieved whether or not a 
conscious observer is involved~\cite{bib_vonneumann}.},
in which case the size of
$A^\prime,A^{\prime\prime},\cdots$ simply depends on the number of particles
in the measurement apparatus. As a matter of
fact, just {\it one} particle living in the same Hilbert space as
$Q$ and $A$ is enough to complete a conceptual measurement, so that
the description of the system $QAA^\prime$ is enough to completely 
model quantum measurement. 
\par

Our model does therefore not fall in the class of
environment-induced decoherence models, simply because information is
not lost to an environment. 
We have a quantum state $Q$ and an ancilla $A$ (which may be composed of
very few degrees of freedom, and does not have to be ``large'').
We suggest that a measurement simply implies
ignoring the quantum system $Q$, which forces the ancilla $A$
to appear in a mixed state. Our model does {\em not}
predict the quantum system $Q$ to be classically correlated with the
ancilla $A$ after the measurement, the cornerstone of standard
environment-induced decoherence models. Rather, we argue that the
classical correlations that emerge from the measurement (by tracing
over $Q$) concern the internal degrees of freedom
of $A$ only. The ancilla is therefore ``self-consistent'', since
arbitrarily dividing $A$ into two halves always provides two
classically correlated subsystems.  In other words, $A$ is never
correlated with $Q$; correlations only appear inside $A$.
Thus, our description appears to be more
fundamental, as it can account for a measurement situation where the
degrees of freedom of $A$ are few and totally controllable (they are
not traced over). In contrast, environment-induced decoherence
models cannot explain the appearance of mixed states
in such systems (see, e.g., \cite{bib_DNA}).
Of course, our model does not preclude a more complex
situation where a macroscopic uncontrollable environment is coupled
with $Q$ and $A$, but we believe such an environment is not conceptually
necessary to interpret a measurement. The apparent irreversibility
(creation of entropy) is traced to the ``hidden'' negative entropy inside
the measured quantum system itself, not to the large environment.
\par

As will be emphasized in Section~\ref{sect_incompatible}, 
the illusion of a wave-function collapse
can be understood by considering consecutive measurements.
A subsequent observation of $Q$ (which is now part of an
entangled system $QAA^\prime$) with another ancilla, say $BB^\prime$,
will result in $BB^\prime$ showing the {\em same} internal correlations
as $AA^\prime$. More importantly, the second ancilla will be
100\% correlated with the first, implying that it reflects the
same exact outcome. This leaves the observer with the illusion that one
definite outcome was recorded by the first ancilla and that any 
subsequent measurement simply confirms that $Q$ is in that state.
In other words, it appears as if the first measurement
projected the quantum state onto an eigenstate, as
reflected by any subsequent measurement. Yet, the only effect of the
measurement on the quantum state is entanglement with the devices, and
all amplitudes of the quantum system are unchanged. Partial observation
of the entangled state leads to all the devices being 100\% correlated.
\par

\subsection{Amplification and reversibility}

\begin{figure}[t]
\caption{Diagrammatic representation of the two-stage unitary measurement.
EPR-entanglement between measured quantum system $Q$ and ancilla $A$ 
(first stage, $U_1$) and entanglement between $QA$ and macroscopic system
$A^\prime A^{\prime\prime}\cdots$ (second stage, observation $U_2$).  
The macroscopic ancilla
$AA^\prime A^{\prime\prime}\cdots$ {\em unconditional} on $Q$ is a mixed
state describing classical correlation. However, $Q$ and 
$AA^\prime A^{\prime\prime}\cdots$ still form an EPR-pair. 
\label{fig3} }
\vskip 0.5cm
\par
\centerline{\psfig{figure=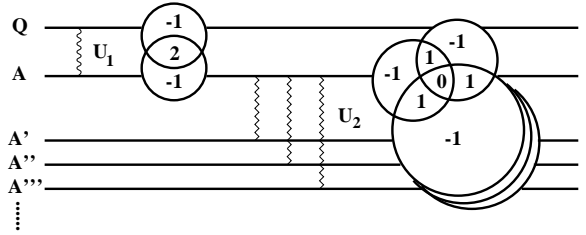,width=3.0in,angle=270}}
\end{figure}

As mentioned above, inducing classical correlations between the quantum
variables $A$ and $A^\prime$ does not lead to a macroscopically observable 
pointer. Rather, the  basic unitary operation  
\begin{equation}
(QA)+A^\prime\stackrel{U}\longrightarrow QAA^\prime \label{unitary}
\end{equation}
must be ``repeated'' $O(10^{23})$ times until a macroscopic
number of quantum particles ($A',A'',\cdots$) are correlated with $A$, such 
that the result can be observed and {\it recorded}. The 
quantum state of the joint system $QAA^\prime A^{\prime\prime}\cdots$ is 
akin to an entangled EPR-nplet with vanishing
entropy. An experimental setup allows the observation of the 
correlations between $A$ and $A^\prime A^{\prime\prime}\cdots$ 
{\em unconditional} on $Q$ (ignoring the quantum state itself), and 
results in
all of the $10^{23}$ particles reproducing (being classically correlated with)
the quantum state of $A$. This process is usually called the amplification, 
or ``classicization'', of the quantum state $A$. The two-stage 
measurement process 
including entanglement and amplification is pictured in Fig.~\ref{fig3}.
\par

Before turning to the question of reversibility, let us stress the fact 
that the creation of entropy (in a subsystem) depends on the initial 
state of $Q$ with respect to the observable under consideration. 
The fact that an {\it arbitrary} state cannot be duplicated (or cloned)
plays a crucial role in the amplification process:
the quantum non-cloning theorem \cite{bib_noncloning}
states that it is possible to amplify a quantum state (e.g.\ the state of $A$)
{\it only} if it belongs to a set of orthogonal states. More precisely,
when a quantum system $Q$ is allowed to interact with an ancilla $A$
in order to measure an observable $O_A$,
the eigenstates $|a\rangle$ of $O_A$ define the set of orthogonal states
that can be amplified (and which lead to a macroscopic device
that reflects the microscopic state). An attempt at amplifying an
{\em arbitrary} quantum state {\em will} generate entanglement between 
the particles constituting the macroscopic object. This entanglement then
is responsible for the generation of randomness in the outcome.
Accordingly, subsystem $(AA^\prime A^{\prime\prime}\cdots)$ carries 
positive {\em unconditional} entropy, 
while the unobserved $Q$ (which is traced over) carries the commensurate 
negative {\em conditional} entropy to allow for the zero entropy pure state 
of the entire entangled system $QAA^\prime A^{\prime\prime}\cdots$.
\par

Let us close this section by stressing that, while quantum measurement 
is conceptually reversible, its irreversible appearance has the same roots
as irreversibility in classical mechanics, as suggested earlier
by Peres~\cite{bib_peres74}. 
As explained previously, the amplification consists in repeating
the basic von Neumann measurement a large number of times, until a
macroscopic number of quantum particles are correlated
with $A$. The {\em whole} (isolated) system is in a {\em pure} entangled state,
but ignoring (tracing over) $Q$ makes the rest of the 
system appear classically correlated.
Yet, {\it no irreversible process} takes place.
Randomness [the probabilities $p_n$ in (\ref{traced})] is generated 
because $A$ already {\it appears}
to be random if one fails to take into account $Q$. This is the measurement 
analogue of the random orientation of half of an EPR pair in an otherwise
fully determined ($S=0$) system. Nothing new
happens by introducing correlations between $A$ and a macroscopic
number of quantum particles ($A^\prime A^{\prime\prime}\cdots$).
However, {\em reversing} the ``observation'' operation [applying a sequence of
inverse unitary transformations of the type (\ref{unitary})]
turns out to be exceedingly difficult in practice. Indeed, one would have 
to reverse {\it every one} of the $O(10^{23})$ unitary operations
that introduced the correlations between the macroscopic set of particles.
While this is possible in principle, it is practically not so 
because missing a single particle that
was involved in the measurement would result in the incorrect unitary (inverse)
transformation, thus failing to restore the initial quantum state. The root
for the practical irreversibility is thus the same for the quantum 
measurement as for the physics of macroscopic classical systems. 
The temporal development is irreversible only because of the practical 
impossibility to control a macroscopic number of initial conditions, while  
the microscopic interactions are all reversible.
\par
As a consequence, we see that only those quantum measurements can be 
reversed for which the ancilla $A$ is {\it not} correlated with a macroscopic 
number of 
particles, i.e., when $A$ is not explicitly observed by a macroscopic 
observer.
However, the reversibility of the first stage of the measurement, the quantum 
entanglement, can, and {\it has been}, achieved. Common lore of
double-slit experiments holds that just providing
the {\it possibility} of performing a measurement (providing the opportunity
to obtain ``which-path'' information, for example) is irreversible. As 
illustrated by the so-called ``quantum-eraser'' 
experiments, this is incorrect~\cite{bib_queraser}.
Indeed, providing the possibility of observation (rather than measurement 
itself) is, according to the unitary quantum measurement theory outlined 
here, just the von Neumann measurement (the first stage, or
EPR entanglement), and is therefore completely reversible. 
In Appendix~\ref{sect_application}, we analyse the quantum eraser
setup within our framework.

\section{Incompatible measurements and uncertainty relations}
\label{sect_incompatible}

We will now show that the uncertainty principle which characterizes the
measurement of two incompatible observables arises naturally
from our unitary description of the measurement process.
We also derive a new bound for the entropic uncertainty relation 
for consecutive measurements which is stronger than the one
in the literature to date.
\par

Let us perform two consecutive measurements on the quantum
system $Q$. First, we measure the observable $O_A$ by allowing
$Q$ to interact with a (first) ancilla $A$. (The amplification stage of the
measurement is ignored here for the sake of simplicity). Subsequently,
we let the system $Q$ interact with an ancilla $B$ in order to
measure observable $O_B$. For illustrative purposes, we assume that
$Q$ is a discrete system which is initially described by the
state vector 
\begin{equation}
|Q\rangle=\sum_{i=1}^N \alpha_i |a_i\rangle
\end{equation}
where $|a_i\rangle$ are the eigenstates of $O_A$ and $N$ is the
dimension of the Hilbert space associated with $Q$ (or $A$, or $B$).
The unitary transformation associated with the measurement of $O_A$
creates an entangled state for the joint system $QA$
\begin{equation}
|QA\rangle=\sum_{i=1}^N \alpha_i |a_i, i\rangle
\end{equation}
where $|i\rangle$ are the basis states of $A$, which label the different
outcomes of the first measurement. In other words, if
$Q$ is in state $|a_i\rangle$, the ancilla $A$ ends up in state
$|i\rangle$. As explained previously, if $Q$ is initially {\em not} 
in one of the
eigenstates of $O_A$, $QA$ will be entangled. Of course, $S(QA)=0$, 
since it evolved unitarily from the pure state $|Q,0 \rangle$.
The marginal density matrix of $A$ is obtained by tracing the density matrix
$\rho_{QA}=|QA\rangle \langle QA|$ over $Q$, yielding
\begin{equation}
\rho_A=\sum_i |\alpha_i|^2 |i\rangle \langle i|\;.
\end{equation}
Consequently, the quantum entropy of $A$ is given by
\begin{equation}
S(A)=H[p_i]
\end{equation}
where $H[p_i]$ denotes the classical (Shannon) entropy
\begin{equation}
H[p_i]= -\sum_ip_i\log p_i 
\end{equation}
associated with the probability distribution $p_i=|\alpha_i|^2$.
This is in perfect agreement with the standard
description of a measurement, which states that the outcome $i$ occurs with
a probability $p_i=|\alpha_i|^2=|\langle a_i|Q \rangle |^2$, i.e., it is 
simply the square of the quantum amplitude $\alpha_i$. Remarkably thus, the 
physical (von Neumann) entropy of $A$ reduces {\em precisely} to  
the Shannon entropy for the outcome of the measurement, which is the one
predicted by standard quantum mechanics. Yet, since $A$ is entangled with
$Q$, the physical entropy of the combined system remains zero.
\par

We now consider the measurement of the second observable $O_B$,
by letting $Q$ interact with $B$. First, we define the unitary operator
$U$ which transforms the eigenstates $|a_i\rangle$ of $O_A$ into
the eigenstates $| b_j \rangle$ of $O_B$: its matrix elements 
are\footnote{This unitary operation is unique up to a 
permutation of eigenstates which is unimportant in this discussion.}
\begin{equation}
U_{ij}= \langle b_j|a_i\rangle\;.
\end{equation}
Obviously, if $O_A$ and $O_B$ commute, $U$ is the identity matrix.
Expressing $|Q\rangle$ in the $|b_j\rangle$ basis and entangling 
it with $B$ in order to measure $O_B$,
we obtain the final state of the system
\begin{equation}
|QAB\rangle=\sum_{i,j=1}^N \alpha_i U_{ij} |b_j, i, j\rangle
\end{equation}
where $|j\rangle$ are the basis states of $B$ (again, this means that
if $B$ is in state $j$ then $Q$ was initially in $b_j$). This is also an 
entangled
state, with zero entropy [$S(QAB)=0$] since it was obtained by evolving
a pure state using two unitary transformations. The marginal density matrix
describing $AB$ (ignoring the system $Q$) is given by
\begin{equation}
\rho_{AB}=\sum_{i,i',j} \alpha_i \alpha^*_{i'} U_{ij}U^*_{i'j}
|i,j\rangle \langle i',j|\;.
\end{equation}
Note that $\rho_{AB}$ {\em cannot} be diagonalized by applying a change
of variable of the product form ($U_A \otimes U_B$), except in the
case where $O_A$ and $O_B$ commute.
The marginal density matrices for $A$ and $B$ are given by
\begin{eqnarray}
\rho_A &=& \sum_i |\alpha_i|^2 |i\rangle \langle i|\;,  \\
\rho_B &=& \sum_{i,j} |\alpha_i|^2 |U_{ij}|^2 |j\rangle \langle j|\;.
\end{eqnarray}
The quantum entropies of $A$ and $B$ then read
\begin{eqnarray}
S(A) &=& H[p_i] \qquad {\rm with}~p_i=|\alpha_i|^2 \;,  \\
S(B) &=& H[q_j] \qquad {\rm with}~q_j=\sum_i p_i q_{j|i}\;.
\end{eqnarray}
where $q_{j|i}=|U_{ij}|^2$ and $H[q_j]$ is the classical (Shannon) entropy
associated with the probability distribution $q_j$.
Here, $q_{j|i}$ can be understood as the {\em conditional} probability
to obtain the outcome $j$ for the second measurement, after having obtained
outcome $i$ for the first one. 
\par
Remarkably, the entropy of the
second measurement $H[q_j]$
is completely compatible with the standard assumption of a collapse
of the wave function in the first measurement. Indeed, it
corresponds exactly to what would be predicted in conventional
quantum mechanics, by
assuming that the wave function was projected on $|a_i\rangle$ with
a probability $p_i=|\alpha_i|^2$ after the first measurement, and
interpreting $|U_{ij}|^2$ as the probability of measuring $j$ on an
eigenstate $|a_i\rangle$ of the first observable.
This reveals how the standard assumption of wave function collapse in 
measurement can be {\it operationally} correct, although we show here 
that it is not the actual physical process. 
Note that the first measurement can be viewed as inducing 
a ``loss of coherence'', as the second measurement yields
$q_j=\sum_i |\alpha_i U_{ij} |^2$ rather than
$q_j=|\sum_i \alpha_i U_{ij} |^2= | \langle b_j | Q \rangle |^2$,
as would be the case if there was no first measurement. For the combined
system $QAB$ on the other hand, there is of course no loss of coherence.
\par

\begin{figure}[t]
\caption{ (a) Ternary entropy diagram for the system $QAB$
(quantum system $Q$, and ancillae $A$ and $B$).
(b) Entropy diagram of the system $AB$ unconditional
on $Q$, describing the sequential measurement of $O_A$ and $O_B$. 
\label{fig4} }
\vskip 0.5cm
\par
\centerline{\psfig{figure=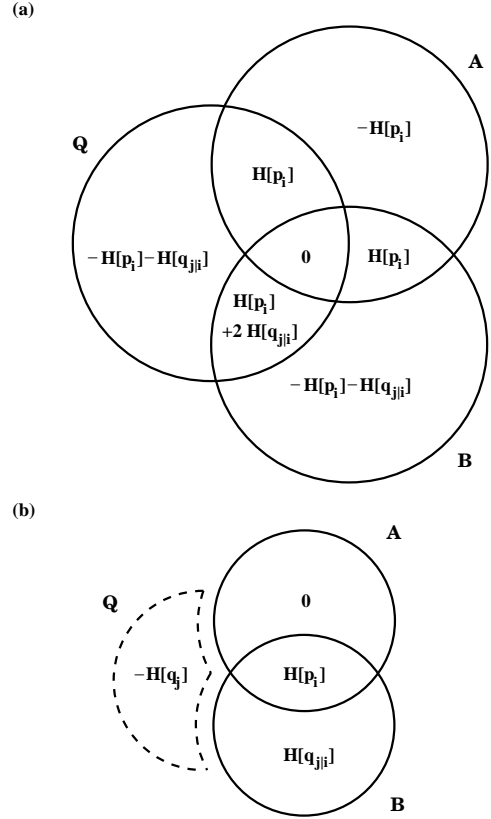,width=2.50in,angle=0}}
\vskip -0.5cm
\end{figure}

The entropy diagram corresponding to the state $QAB$ is shown
in Figure~\ref{fig4}a. The entropy of $A$ (resulting from the 
first measurement)
is  $S(A)=H[p_i]$, whereas the entropy of $B$ (resulting from the
second measurement) is 
\begin{equation}
S(B)=H[q_j]=H[p_i]+H[q_{j|i}]\;, 
\end{equation} where we defined the (classical) {\em conditional} entropy
\begin{equation} 
H[q_{j|i}]=- \sum_{i,j} p_i q_{j|i} \log q_{j|i}\;.
\end{equation}
This last quantity
represents the {\em additional} amount of entropy that appears due to the
second measurement. Figure~\ref{fig4}b depicts the apparent
entropy diagram of $AB$ unconditional on $Q$, illustrating the basic equation
\begin{equation}
S(A)+S(B|A)=H[p_i]+H[q_{j|i}]=H[q_j] \label{uncertainty}
\end{equation}
relating the entropy of the first and the second measurement.
Note that, despite the asymmetry between $A$ and $B$ ($O_A$ is measured
first), Eq.~(\ref{uncertainty}) can be rewritten in symmetric form
\begin{equation}
S(A)+S(B) \ge H[q_j] \label{uncert}
\end{equation}
since the mutual entropy $S(A{\rm :}B)$ is always positive.
Equation (\ref{uncertainty}) plays the role of
an uncertainty relation for entropies, expressing the fact the
the sum of the entropies resulting from the measurement of $O_A$ and
$O_B$ is constant. If we were to try to reduce the entropy associated with
one of them, then the other entropy would increase.
In order to have a genuine ``entropic uncertainty relation''
for consecutive measurements, independent
of the initial state of $Q$, it is necessary to minimize
the right-hand side of Eq. (\ref{uncertainty}) over $|Q\rangle$
(i.e., over the $\alpha_i$'s). The convexity of Shannon entropy implies 
that $H[q_j]$ is minimized in the case where the $p_i$ distribution is
maximally peaked, that is, when the initial state of $Q$ is an
eigenstate $|a_i\rangle$ of the first observable. 
In this case, $S(A{\rm :}B)=H[p_i]=0$, and therefore, assuming 
$|Q\rangle=|a_i\rangle$ (for instance) yields
\begin{equation}
S(A)+S(B) \ge H[q_{j|i}]_{i~{\rm fixed}}
  \equiv  -\sum_j q_{j|i} \log q_{j|i}
\end{equation}
Then, minimizing over $i$,
we obtain the entropic uncertainty relation
\begin{equation}
S(A)+S(B) \ge \min_i H[q_{j|i}]_{i~{\rm fixed}} 
= \min_i H[|U_{ij}|^2]_{i~{\rm fixed}}   \label{entropic}
\end{equation}
Physically, this means that the sum of the entropies is bounded from
below by the Shannon entropy corresponding to the expansion
of an eigenstate of $O_A$ into the basis of eigenstates of $O_B$
(more precisely, the eigenstate which minimizes the Shannon entropy).
Note that our entropic uncertainty relation (\ref{entropic}) is stronger than
the Deutsch-Kraus exclusion principle~\cite{bib_deutsch,bib_kraus,bib_uffink},
which states that
\begin{equation}
S(A)+S(B) \ge - \log c
\end{equation}
where $c=\max_{i,j} |U_{ij}|^2$.
Indeed, it is easy to see that 
$\min_i H[|U_{ij}|^2]_{i~{\rm fixed}} \ge - \log c$.
\par

In the case of complementary observables (i.e., if the distribution
of $O_A$ values is uniform for any eigenstate of $O_B$ and vice versa),
one obtains the simple entropic uncertainty relation~\cite{bib_kraus,bib_hall}
\begin{equation}
S(A)+S(B) \ge \log N
\end{equation}
where $N$ is the dimension of the Hilbert space, as expected. This bound just
corresponds to the situation where the conditional entropy $S(Q|AB)$ takes on
the largest negative value compatible with the dimension of the Hilbert space
of $Q$. This is for instance the case if one measures any two 
spin-projections of a spin-1/2 particle. In this case, we obtain 
\begin{equation}
S(\sigma_x)+S(\sigma_y) \ge 1\;.
\end{equation}
For the case of two commuting observables ($[O_A,O_B]=0$),
we find $U_{ij}=\delta_{i,j}$ and therefore
$S(A)+S(B) \ge 0$, reflecting that they can be
measured simultaneously with arbitrarily high accuracy.
In situations that are intermediate between compatible and
complementary (maximally incompatible) observables, our
bound is demonstrably more constraining than the one of Deutsch and Kraus.
Let us show this for the simple case of a two-dimensional Hilbert space.
\par

\begin{figure}[t]
\caption{ Lower bound for the entropic uncertainty relation
in a spin-1/2 Hilbert space. The solid line represents our bound
[Eq.~(\ref{eq_ourbound})], while the dashed line stands for
the Deutsch-Kraus bound [Eq.~(\ref{eq_deutsch})], for $\theta$ between
0 and $\pi/2$. \label{fig_uncert} }
\vskip 0.5cm
\par
\centerline{\psfig{figure=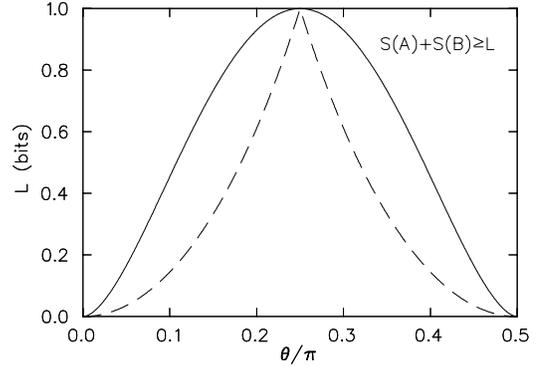,width=2.70in,angle=90}}
\vskip -0.5cm
\end{figure}

For a general $2\times 2$ unitary matrix $U_{ij}$, with
$|U_{11}|^2=|U_{22}|^2=\cos^2 \theta$, 
$|U_{12}|^2=|U_{21}|^2=\sin^2 \theta$, and $\theta$
an angle parameter, the Deutsch-Kraus uncertainty relation is 
\begin{equation} \label{eq_deutsch}
S(A)+S(B) \ge - \log \max \left\{ \cos^2 \theta , \sin^2 \theta \right\} \;,
\end{equation}
whereas we find 
\begin{equation} \label{eq_ourbound}
S(A)+S(B) \ge H \left[ \cos^2 \theta, \sin^2 \theta \right]  \;.
\end{equation}
In Fig.~\ref{fig_uncert}, we compare the right-hand sides of 
Eqs.~(\ref{eq_deutsch}) and (\ref{eq_ourbound}), illustrating that
the bounds are equal only for completely compatible ($\theta=0,\pi/2$) 
or maximally incompatible ($\theta=\pi/4$) observables.

\section {Interpretation}
\label{sect_interpretation}

In this section we comment on the implications of unitary
quantum measurement and the concept of quantum entanglement for
the foundations and the interpretation of quantum mechanics.
\par
The inability to consistently describe the measurement process
in quantum mechanics--the quantum measurement paradox--has
seriously discredited
the foundations of a theory that otherwise describes the 
microscopic world succinctly, effortlessly, and correctly. The 
questions that we would like to address anew here concern the 
relation between quantum and classical concepts, the Schr\"odinger 
``cat paradox'', as well as the interpretation of the complementarity 
principle.  
\par
In standard quantum mechanics, the criterion to 
decide whether a classical or a quantum picture is more adequate
generally involves comparing a representative unit of action of the system 
under consideration $S_{\rm typ}$ with the unit $\hbar$.
 Such a criterion suggests that any macroscopic system that fulfills 
$S_{\rm typ}\gg\hbar$ behaves classically.
Yet, the present paper proposes that EPR-entangled systems, whether 
microscopic or macroscopic, are fundamentally
quantum and can in {\em no limit} be understood classically. 
We would like to suggest here that a degree of
freedom {\em appears} classical if it is composed of many 
[$O(10^{23})$] classically correlated internal variables. This occurs 
precisely when part of an entire isolated system 
which is in a pure quantum state is ignored (i.e., unobserved and 
traced over). Note that tracing over just {\em one} degree of freedom 
that is entangled is enough to promote the classical appearance!
Tracing over an unobserved degree of freedom is {\it not} a physical 
process, and is thus not described by any
time evolution. 
Rather, a quantum measurement forces the observation of correlations
{\em unconditional} on part of a (quantum inseparable) system. 
Thus, any classical degree
of freedom has a ``classical appearance'' only because it is part
of a larger quantum inseparable system in a pure state.
\par

Let us consider this in more detail, as it suggests a very simple and
satisfying explanation for the Schr\"odinger cat paradox. In this, perhaps 
the most well-known and most puzzling of all {\em gedankenexperimente},
the first stage of the measurement concerns a 
decaying atom and its emitted particle (say, a photon). 
Let us assume, as is usual, that the wavefunction (after some time) 
is a superposition of an ``excited'' atom $A^\star$ and the vacuum,
and a decayed atom $A$ with one photon:
\begin{equation}
|\Psi_0\rangle=\frac1{\sqrt2}\bigl(|A^\star,0\rangle + |A,1\rangle\bigr)\;,
\end{equation}
i.e., both atom and photon form an entangled state with vanishing 
overall entropy. 
Then, in the second stage of the measurement,
the $O(10^{23})$ atoms forming the cat interact with the photon,
forming an EPR-nplet for the entire quantum
state -- of course still a pure state. If we simplify
the problem by assuming that the cat's quantum variable is dichotomic, with
live and dead cat eigenstates $|L\rangle$ and 
$|D\rangle$, the wave function becomes 
\begin{equation}
|\Psi_1\rangle=\frac1{\sqrt2}\biggl(|A^\star,0,L\rangle + 
|A,1,D\rangle\biggr)\;.
\end{equation}
Tracing over the initial atom (the experiment after all involves 
monitoring the cat, not the atom),
one obtains a mixed state where all the $10^{23}$ atoms are {\em correlated}
with the emitted particle, i.e., they are arranged in such a way that the
cat is either dead or alive (with probabilities 1/2):
\begin{equation}
\rho_{\gamma,\rm cat}= \frac12\biggl(|0,L\rangle\langle 0,L| + 
|1,D\rangle\langle 1,D|\biggr)\;.
\end{equation}
This macroscopic system has an entropy of 1 bit, that
is, randomness has been created. More importantly, the density
matrix is equal to that of a statistical ensemble prepared
with equal numbers of dead and living cats, making both situations
(the experiment and the preparation) physically {\em indistinguishable}.
The randomness created in the cat-$\gamma$ subsystem is compensated by a 
conditional entropy of --1 bit for the decaying atom. 
Since the entire system has vanishing entropy, it is still completely 
determined. Moreover,
no such thing as a collapse of the cat wave function happens when 
the box is opened to an observer; what happens is simply that now all the atoms
of the observer become also entangled with those of the cat:
\begin{equation}
|\Psi_2\rangle=\frac1{\sqrt2}\biggl(|A^\star,0,L,l\rangle + 
|A,1,D,d\rangle\biggr)\;.
\end{equation}
where we introduced the dichotomic observer states $|l\rangle$ and $|d\rangle$
describing the observation of the live or dead cat. The corresponding
marginal density matrix is
\begin{equation}
\rho_{\gamma,\rm cat, obs}= \frac12\biggl(|0,L,l\rangle\langle 0,L,l| + 
|1,D,d\rangle\langle 1,D,d|\biggr)\;.
\end{equation}
The observer notices that the cat is either dead or alive, and thus the 
observer's own 
state becomes classically correlated with that of the cat, although,
in reality, the entire system (including the atom, the $\gamma$, the cat, and 
the observer) is in a {\em pure entangled} state. It is {\em practically} 
impossible, although not in principle, to undo this observation, i.e., 
to resuscitate the cat, or, more precisely, to come back to the initial
decaying atom, with a living cat and an ignorant observer
\begin{equation}
|\Psi_2\rangle\stackrel{U_2^{-1}}{\longrightarrow} | \Psi_1 \rangle
\stackrel{U_1^{-1}}{\longrightarrow} | \Psi_0 \rangle \;,
\end{equation}
since it requires to enact the inverse unitary transformations on all the
atoms forming the observer and the cat. This  
irreversibility is completely equivalent to the irreversibility in
classical mechanics. Indeed, classically, to reverse the microscopic 
time evolution, it is necessary to invert the velocity of all the particles, 
the practical impossibility of which gives a
macroscopic irreversible aspect to time evolution. In quantum
mechanics, it is necessary to undo any unitary evolution associated
with all interactions that particles have undergone, so that reversibility
is practically impossible if a macroscopic number of particles have
been involved. We are led to conclude that irreversibility is {\em not} an 
inherent feature of quantum mechanics.
\par

Finally, the present approach sheds light on the origins of the complementarity
principle, or wave-particle duality. On the one hand, we see that the wave 
function {\em completely} describes a quantum state, a fact eloquently
argued for by Bohr. On the other hand, we cannot escape the appearance
of randomness in quantum measurement. These facts were interpreted
by Bohr to be ``complementary'' to each other,
much as the wave nature of quantum objects was viewed as ``complementary''
to its particle nature. Our identification of von Neumann entropy as the 
real, physical, entropy of a system corroborates that the quantum
wave function does indeed provide a complete description of the quantum state,
since the von Neumann entropy of a pure state is zero. Yet, we find that 
randomness is not an essential cornerstone of quantum measurement, but 
rather an illusion created by it. Thus, we are led to conclude that 
complementarity is a working concept, but has no ontological basis as a 
principle. The same appears to be true for the wave-particle duality. 
On the one hand we agree that quantum systems, due to the superposition
principle, are wave-like in nature. This is inherent in the 
``completeness postulate of the density matrix'' (see, e.g., \cite{bib_peres}),
which implies that two systems prepared in the same density matrix,
but by making different mixtures of pure states, are completely 
{\it indistinguishable}. On the other hand, the particle aspect
of a quantum object emerges simply from the measurement process, when a 
wavefunction interacting with a measurement device appears as a mixed state.
Thus, as we unmask the particle-like behavior of quantum systems to be an
{\em illusion} created by the incomplete observation of a quantum (entangled)
system with a macroscopic number of degrees of freedom, we are led to conclude
once more that the wave-nature of quantum systems is {\em fundamental}, and 
that there is no particle-wave duality, only an apparent one. 

\section{Conclusion}
\label{sect_conclusion}

In conclusion, we are able to reconcile unitary evolution of quantum 
states and the apparent creation of randomness in a minimal model
of the measurement process.
This is achieved via the introduction of an elementary quantum measurement 
process (the EPR entanglement) in which entropy is conserved
by balancing randomness with negative entropy. 
We show how the usual probabilistic 
results of quantum mechanics arise naturally in this description,
paving the way for a fully consistent description of quantum mechanics
in which the measurement device is {\em not} accorded a privileged role. 
This description does not require the concept of wave function collapse 
or the presence of a macroscopic environment
in order to predict the results of quantum experiments, thereby removing
the special status of quantum mechanics as far as irreversibility is 
concerned. 
In addition, our analysis shows that, in spite of its appearance,
any classical system is
in fact an entity which is part of a larger quantum system. We believe 
this answers the question about the location of the frontier between the 
quantum and the classical world, with respect to measurement. We answer 
that there is no classical world, only the classical appearance of 
part of a quantum world. 
This view is especially satisfying as measurement, bereft of its
special status outside of quantum mechanics (which it had been accorded to
by the Copenhagen interpretation) and unencumbered by external notions
such as consciousness (as advocated by von Neumann) is now 
part of a consistent theory defined without recourse to classical notions
which, after all, should appear as a limit of a quantum theory only. 
\par

\acknowledgements
We would like to express our deepest gratitude to Prof. Hans Bethe, who after 
reading Ref. \cite{bib_neginfo} informed us that ``negative [...] entropy
solves the problem of the reduction of the wave packet'' and encouraged us
to pursue this avenue. We are also grateful for many 
enlightening discussions on this topic while he was visiting the Kellogg 
Lab in the spring of 1996. Finally, we thank A. Peres and M.J.W. Hall
for useful comments on the manuscript.
This work was supported in part by NSF Grant
Nos. PHY 94-12818 and PHY 94-20470.

\appendix

\section{Standard quantum experiments}
\label{sect_application}

In this appendix we apply our quantum measurement theory to 
standard experiments, in order to illustrate how the usual 
quantum probabilistic results emerge in a unitary treatment.

\subsection {Stern-Gerlach experiment}

\begin{figure}[t]
\caption{ (a) Setup of the Stern-Gerlach experiment.
(b) ``Consistency'' requirement for two sequential
Stern-Gerlach experiments illustrating the appearance of 
classical correlation.  \label{fig5} }
\vskip 0.5cm
\par
\centerline{\psfig{figure=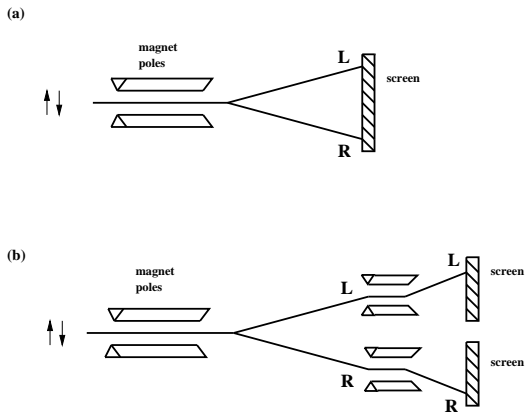,width=2.70in,angle=0}}
\end{figure}

In the Stern-Gerlach experiment, a beam of atoms is guided through
an inhomogeneous magnetic field $B_z$ normal to the direction of motion
of the atoms (see Fig.~\ref{fig5}a). In this field, the atoms experience a
force deflecting them out of the beam, depending on the orientation
of their magnetic moments with respect to the magnetic field axis.
The beams are collected a distance away on a screen. Let us assume here
for simplicity that the magnetic moments of the atoms take on only two
different values ($s=1/2$), and define $\sigma_z$ eigenstates 
$|\uparrow\rangle$ and $|\downarrow\rangle$. If the incident beam 
consists out of atoms prepared in a $\sigma_x$ (say) eigenstate, the
initial state is a quantum superposition
\begin{equation}
|\Psi_{\rm beam}\rangle=\frac1{\sqrt2}\bigl(|\uparrow\rangle+
|\downarrow\rangle\bigr)\;.
\end{equation}
The auxiliary variable, or ancilla, is in this case a spatial
location, say left or right ($L,R$). Applying the magnetic field 
then completes the von Neumann measurement
\begin{equation}
|\Psi\rangle=\frac1{\sqrt2}\bigl(|\uparrow,L\rangle+
|\downarrow,R\rangle\bigr)\;.
\end{equation}
Through this operation, the different spin-orientations have been ``tagged''
(the $\uparrow$ spin is tagged with a left location, and
conversely),
but it is incorrect to assume that spin-orientations and locations
are now {\em correlated}. Much more than that, they are {\em entangled}:
locations and spin-orientations form EPR pairs. The second stage of
measurement (amplification) occurs on the screen. Collecting the 
particles ignores the spin-orientation entirely such that the 
particles of the screen become classically correlated with the 
location variable, forming a type II classically correlated system
carrying one bit of entropy. Let us emphasize here that the measurement
of the location variable $(L,R)$ does {\em not} allow us to infer the
spin orientation of the atom. Thus, even though the particle beam
was deflected in the $z$-direction (as if the beam was composed of
atoms with magnetic moments quantized in the $z$-direction), such a classical
description is misleading.
\par

Denoting as usual the system (atom) with $Q$,
the ancilla (location) with $A$ and the screen with $A^\prime$ (with 
eigenstates $|l\rangle$ and $|r\rangle$), we obtain
\begin{equation}
\rho_{AA^\prime}={\rm Tr}_Q(\rho_{QAA^\prime}) =
\frac12\biggl(|L,l\rangle\langle L,l|
+|R,r\rangle\langle R,r|\biggr)\;
\end{equation}
which is the standard  result: the spot on the screen reflects the 
$L-R$ variable (classical correlation). Yet, the entropy of the 
combined system $QAA^\prime$ has not changed, still being zero. 
The randomness in the measurement result
(the bit of entropy in the $AA^\prime$ system) 
is cancelled by the negative entropy of the unobserved quantum state $Q$, 
\begin{equation}
S(Q|AA^\prime)=-1\;.
\end{equation}
It is important to observe that the randomness which may appear in the
measurement of the position (collecting the particles on
a screen or a detector) does {\it not} occur because there were unknown
internal degrees of freedom, which along
with the wave function, would be needed to completely describe the particle
(cf. hidden-variable theory). The wave
function {\it entirely} defines the state (it is indeed of zero entropy). 
\par

It is well-known that if 
a second magnetic field gradient is used in order to perform a second
Stern-Gerlach measurement (foregoing the collection on the screen) as depicted
in Fig.~\ref{fig5}b, one obtains two correlated variables:
the position $x$ after the first, and $y$ after
the second field gradient.
The standard interpretation is that, once the wave function has been projected
(via the first field gradient), only positive (negative) spin-projection 
particles are left in the $L(R)$ beam, so that the second measurement 
is incapable of splitting the beam again.
This is a basic requirement
for consecutive measurements of the same observable on a quantum system.
In reality, this is nothing else than the classical correlation
which appears when a pure quantum state is observed only partially. The two
position variables $x$ and $y$ are classically correlated (mixed state)
since one is ignoring the spin orientation ($x, y,\sigma_z$ form an 
EPR-triplet).
This experiment is practically irreversible since the second stage of the
measurement (classicization) occurs when {\it detecting} the particle 
after the second field gradient. Whenever no detector is placed
after the field gradient, the ``measurement'' is easily reversible, but 
in that case it has not been observed by a macroscopic observer.
\par

\subsection{Quantum eraser}

The quantum eraser experiment (see \cite{bib_queraser}) provides
a nice demonstration of how the first stage (von Neumann measurement,  
or ``tagging'') can be reversed. Several versions of this experiment have 
been performed. However, we restrict ourselves here to an idealized such 
experiment for convenience.
\par

\begin{figure}[t]
\caption{ Setup for the ``quantum eraser'' in the two-slit experiment. The
detector in front of the eraser is not part of the standard setup, and
illustrates the impossibility of storing the information before erasure. 
 \label{fig6} }
\vskip 1.0cm
\par
\centerline{\psfig{figure=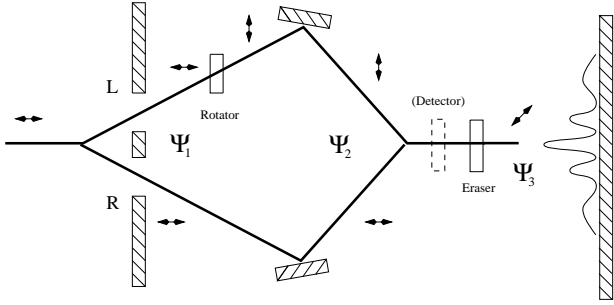,width=3.20in,angle=270}}
\end{figure}

An eraser experiment can be visualized as a two-slit experiment using
a beam of horizontally polarized photons (see Fig.~\ref{fig6}). This 
beam is subsequently split in a crystal. When the split beams
recombine, they produce the well-known interference pattern. 
However, a polarization
rotator placed on, say,  the left path (so that the polarization of one
of the split beams--the left one--is changed from horizontal
to vertical) will cause the interference pattern to vanish. This is
in agreement with Feynman's rule: the paths are
distinguishable since a photon traveling via the left path is
vertically polarized at the screen, while a photon traveling along
the right one remains horizontal. The standard explanation is
that providing the ``which-path'' information precludes the existence
of interference. The quantum eraser idea is that this which-path 
information can be erased, by inserting 
a polarization filter aligned on the {\em diagonal} direction
between the {\em recombined} beams and the screen. Such a 
procedure makes it impossible
to tell whether a photon was horizontally or vertically polarized beforehand.
Accordingly, the interference pattern on the screen is resurrected.
\par

We start with a pure beam of horizontally polarized photons 
(see Fig.~\ref{fig6}). 
After the splitting of
the beam, the quantum state of the photon is described by the state vector
\begin{equation}
|\Psi_1\rangle= \frac1{\sqrt2}\biggl(|L\rangle +|R\rangle\biggr) |H\rangle \;,
\end{equation}
a function of two dichotomic variables: a location variable
$\lambda=L$ (left) or $R$ (right), and a polarization variable
$\sigma=H$ (horizontal) or $V$ (vertical). This describes a  
superposition of a left-photon and a right-photon after 
the splitting of the beams. The polarization
rotator placed on the left path represents the first stage of the
measurement: it can be viewed as a ``tagging'' operation 
(the left path is tagged with a vertically polarized photon and conversely)
resulting in the state
\begin{equation}
|\Psi_2\rangle= \frac1{\sqrt2}\biggl(|L,V\rangle+|R,H\rangle\biggr)\;.
\label{psi2}
\end{equation}
The crucial point is that, after tagging,
the location and polarization variables are {\it entangled} 
and form an EPR-pair. Assuming, as is usually done,
that the photon is either on the left path (with a vertical
polarization) or on the right path (with a horizontal polarization) 
is classical intuition but decidedly {\it wrong}. 
We cannot witness {\em classical} correlation between location ($L$ or $R$)
and polarization ($H$ or $V$); rather, the variables are {\it entangled} (or
super-correlated) carrying negative conditional entropies ensuring that the 
total entropy vanishes.  
Indeed, the state $|\Psi_2\rangle$ is still a pure state, since it evolved
from $|\Psi_1\rangle$ by a unitary transformation. At this stage,
measuring the location $\lambda$ of the photon (ignoring its polarization 
$\sigma$)
yields a random variable (ignoring half of the EPR-pair gives a mixed state
with positive entropy). Equivalently, measuring the polarization $\sigma$ 
of the photon after recombining the beams (ignoring the phase hidden in
the location variable $\lambda$)
also yields a random variable. However, in both cases, this positive entropy
is exactly compensated by a negative conditional entropy such as to preserve
an overall vanishing entropy. Location and polarization play the role of
conjugate (or incompatible) variables that cannot be measured simultaneously.
The entanglement in $|\Psi_2\rangle$ is responsible for the loss of coherence
in the location variable (the marginal density matrix of $\lambda$ is a 
mixture)
which results in the disappearance of the interference pattern.
This is obvious since the cross-terms in the square of
$|\Psi_2\rangle$ vanish because $|V\rangle$ and $|H\rangle$ are orthogonal.
\par

Yet, it can be seen easily that the eraser (the diagonally oriented
polarization filter placed in front of the screen) reverses 
the ``tagging'' operation,
so that the quantum state $|\Psi_2\rangle$ evolves 
back to a pure state 
\begin{equation}
|\Psi_3\rangle=\frac1{2\sqrt2}\biggl( |L\rangle+|R\rangle \biggr)
\biggl( |H\rangle+|V\rangle \biggr) 
\end{equation}
proportional to $|\Psi_1\rangle$, up to a 
trivial rotation of the polarization vector.
This resuscitates the interference pattern as the location variable is
now {\em unentangled}. Indeed, the square of the wavefunction at position
$x$ on the screen is
\begin{equation}
|\Psi_3|^2 = \frac14\biggl(|\psi_L(x)|^2+|\psi_R(x)|^2
+2{\rm Re}\left[\psi^\star_L(x)\psi_R(x)\right]\biggr)\;
\end{equation}
where $\psi_L(x)=\langle x|L\rangle$ for example. 
The quantum
eraser experiment only concerns the first stage of the measurement,
that is the {\it possibility} of observing a measurement.
As explained earlier, only the latter can be reversed in practice, whereas
the macroscopic recording of the polarization is (practically) irreversible.
An attempt at recording the polarization $\sigma$
of the photon after recombination but {\em before} erasure
(see Fig.~\ref{fig6}) to cheat the eraser into delivering
an interference pattern {\em and} which-path information,
involves {\em entangling} the polarization with an ancilla $A$ with eigenstates
$|h\rangle$ and $|v\rangle$:
\begin{equation}
|\Psi_2^\prime\rangle = 
\frac1{\sqrt2}\biggl(|L,V,v\rangle+|R,H,h\rangle\biggr)\;.
\end{equation}
Such an action is enough to thwart any
attempt at recovering the interference pattern. 
Indeed, the action of the eraser on $|\Psi_2^\prime\rangle$ yields
\begin{equation}
|\Psi_3^\prime\rangle=\frac1{2\sqrt2}\biggl( |L,v\rangle+|R,h\rangle \biggr)
\biggl( |H\rangle+|V\rangle \biggr)\;, 
\end{equation}
leaving the location variable $\lambda$ {\em entangled} with $A$ (which is 
typically
a macroscopic number of internal variables which are classically
correlated when ignoring $\lambda$).
In contrast with $|\Psi_3\rangle$, $|\Psi_3^\prime\rangle$ does not give
rise to an interference pattern, as it is completely analogous
to Eq.~(\ref{psi2}). 
\par
The present discussion illustrates Feynman's rule stating that,
in the case of a double-slit experiment,
a quantum state behaves as a particle whenever which-path information is
extracted, and as a wave otherwise. As we saw above, which-path information 
is obtained by entangling the location variable $\lambda$. This
operation by itself generates the appearance of a mixed state (and
commensurate particle-like behavior) from a pure state (with wave-like
behavior).

\end{document}